\RequirePackage{lineno}
\documentclass[nofootinbib,superscriptaddress,twocolumn,showpacs,amsmath,prl,preprintnumbers]{revtex4}
\usepackage{graphicx}
\usepackage{epsfig}
\usepackage{amssymb}
\usepackage{changes}
\usepackage{lineno}

\def\Vec#1{\mbox{\boldmath $#1$}}
\def\beq{\begin{equation}}
\def\eeq{\end{equation}}

\def\beqy{\begin{eqnarray}}
\def\eeqy{\end{eqnarray}}

\newcommand{\pA}{\mbox{$p$A}}
\usepackage{color}

\begin{document}
\title{Evidence for $x$-dependent proton color fluctuations in
  \pA\ collisions at the LHC}
 \author{M. Alvioli}
\affiliation{Consiglio Nazionale delle Ricerche, Istituto di Ricerca per la Protezione Idrogeologica,
  via Madonna Alta 126, I-06128 Perugia, Italy}
  \author{B.A.~Cole}
\affiliation{Physics Department, Columbia University, New York, NY 10027, USA}
 \author{L. Frankfurt }
\affiliation{Department of Particle Physics, School of Physics and Astronomy, Tel Aviv University, Tel Aviv,  Israel}
\author{D.V.~Perepelitsa}
\affiliation{Brookhaven National Laboratory, Upton, NY 11973, USA}
\author{M. Strikman}
\affiliation{104 Davey Lab, The Pennsylvania State University,
  University Park, PA 16803, USA}
\date{\today}

\begin{abstract}
The centrality-dependence of forward jet production in \pA\ collisions
at the LHC has been found to grossly violate the Glauber model
prediction in a way that depends on the $x$ in the proton. We argue
that this modification pattern provides the first experimental
evidence for $x$-dependent proton color fluctuation effects.
On average, parton configurations in the projectile proton containing
a parton with large $x$ interact with a nuclear target with a
significantly smaller than average cross section and have smaller than
average size.  
We implement the effects of fluctuations of the interaction strength and, using the ATLAS analysis of how hadron production at backward rapidities depends on the number of wounded nucleons, make quantitative predictions   for the centrality
dependence of the jet production rate as a function of the
$x$-dependent interaction strength $\sigma(x)$. We find that
$\sigma(x)\sim 0.6 ~\left<\sigma\right>$ gives a good description of
the data at $x=0.6 $. These findings support an explanation of the
EMC effect as arising from the suppression of small size nucleon
configurations in the nucleus.
\end{abstract}
\maketitle


Studies of microscopic nucleon structure have progressed from probing
single parton distributions to the study of generalized parton
distributions (3D single parton distributions) and of parton--parton
correlations in multi-parton interactions. Here we argue that
correlations between soft and hard scattering processes in very high
energy proton--nucleus (\pA) collisions probe how the transverse area
occupied by partons in a fast nucleon depends on the light cone
momentum fraction $(x)$ of the trigger parton. 
The projectile nucleon propagates through the nucleus in a
frozen quark--gluon configuration, leading to the coherence of its
interactions. Due to the color screening property of high energy QCD,
the overall interaction strength of a color neutral configuration
drops with a decrease in its transverse size~\cite{CT}. Such 
phenomena have been observed directly in, for example, 
hard diffractive jet production~\cite{Aitala:2000hc} and in exclusive meson
production \cite{Clasie:2007aa,ElFassi:2012nr}. The presence of configurations
of the nucleon which have smaller than average interaction strength results in
new phenomena which are observable in the centrality dependence of
single jet~\cite{ATLAS:2014cpa} or dijet~\cite{Chatrchyan:2014hqa} production
in \pA\ collisions at the Large Hadron Collider (LHC).

To visualize the origin of fluctuation phenomena in high energy
processes, consider the propagation of an ultra-relativistic
positronium atom through a slab of matter with density $\rho$ and
length $L$. From the uncertainty principle and Lorentz dilation of the
interaction, the transverse distance $r_{t}$ between the electron and
positron remains constant over a longitudinal coherence distance
$l_{coh} \sim (1/\Delta E) \sim \frac{2P_{pos}}{M^2-m_{pos}^2}$, where
$P_{pos}$ is the momentum of the positronium and $M$ is the mass of
the intermediate (diffractive) state. In the limit $l_{coh} \gg L$, a
new absorption pattern emerges. This is due to coherence effects and
to the dependence of the interaction cross-section $\sigma$ of the
$e^+e^-$ dipole with a target on $r_{t}$, which at small transverse
distance is $\sigma(r^2_t) \propto r^2_t$.

In this limit, the probability of an inelastic interaction, $1-\kappa$,
is obtained by summing over a complete set of diffractive intermediate
states with different $r_t$, weighted by the value of the positronium
wave function at $r_t$,
\begin{equation}
  \kappa=\int{dz \, d^2r_t} \; \psi^2 (z,r_t)  \, \exp
  \left[-\sigma_{in}(r^2_t) \rho L\right].
  \label{eq1}
\end{equation} 
\noindent
Here, $\psi(z,r_t)$ is the positronium wave function normalized as
$\int \psi^2(r)d^3r=1$, and $\sigma_{in}$ is the inelastic
positronium--target cross-section.  Then, $\kappa$ is the probability
for positronium to transform to an $e^+e^-$ pair without inelastic
interactions.  For $\sigma_{in} \rho L\gg 1$, the survival rate is
$\kappa \approx 2/(\sigma_{in}\rho L)$ \cite{Frankfurt:1991rk}, which
is larger than the naive expectation $\approx\exp (-\sigma_{in} \rho
L)$. Since $\sigma_{in}$ depends on $r_t$, positronium can be captured
in a larger (smaller) configuration by selecting events with more
(fewer) excited atoms in the target.

In QCD, fluctuations in the interaction strength of a hadron $h$ with
a nucleon originate from fluctuations in both the transverse size and
in the number of constituents of the hadron. We refer to both
generically as color fluctuations (CF).  CF effects can be accounted
for by introducing a probability distribution, $P_h(\sigma)$, for the
hadron to be found in a configuration with total cross-section
$\sigma$ for the interaction. This probability distribution obeys the sum rules $\int
P_h(\sigma)d\sigma=1$ and $\int P_h(\sigma)\sigma
d\sigma=\left<\sigma\right>\equiv \sigma_{tot}^{hN}$ where $\left<\sigma\right>$
is the configuration-averaged (total) cross section. The variance of the
distribution divided by the mean squared,  $\omega_\sigma$, is given by the optical
theorem~\cite{Miettinen:1978jb,Good:1960ba},

\begin{equation}
\omega_{\sigma}= (\left<\sigma^2\right>/\left<\sigma\right>^2 -1)= 
\left.{{d\sigma(h+ p\to X+p) \over dt}\over { d\sigma(h+ p\to h+p) \over dt}}\right\vert _{t=0},
\label{diffr}
\end{equation}

\noindent where a sum over diffractively produced states $X$,
including the triple Pomeron contribution \cite{Alvioli:2014sba}, is
implied.  Fixed target and collider data~\cite{Blaettel:1993rd} indicate that
$\omega_{\sigma}$ for the proton first grows with energy, reaching
$\omega_{\sigma} \sim 0.3$ for $\sqrt{s} \sim 100$~GeV, then decreases
at higher energies to $\omega_{\sigma} \sim 0.1 $ at the
LHC~\cite{Alvioli:2014sba}.

Several considerations constrain the shape of
$P_h(\sigma)$~\cite{Blaettel:1993rd}.  For values 
of $\sigma \sim \langle
\sigma \rangle$, $P_h(\sigma)$ is expected to be Gaussian due to small
fluctuations in the number of, and in the transverse area occupied by,
partons. This expectation is supported by an
analysis~\cite{Blaettel:1993rd} of coherent diffraction measurements
in proton--deuteron collisions~\cite{Akimov:1976hg}.  For $\sigma \ll
\left<\sigma\right>$, configurations with a small number constituents,
$n_q$, localized in a small transverse area should dominate, leading
to $P_h(\sigma)\propto \sigma^{n_q-2}$~\cite{Blaettel:1993rd}. For
protons, the resulting form of $P_p(\sigma)$ and values of $\rho,\sigma_0,\Omega$
were chosen to smoothly interpolate between both regimes while reproducing
measurements of the first three moments of the distribution. It is given by
\begin{eqnarray}
P_p(\sigma)\,=\frac{\rho}{\sigma_0}\left(\frac{\sigma}{\sigma\,+\,\sigma_0}\right)
\,\mathrm{exp} \left\{-\frac{(\sigma/\sigma_0\,-\,1)^2} {\Omega^2}\right\}\, .
\label{psigma}
\end{eqnarray}
For the Gaussian distribution $\omega_\sigma=\Omega^2/2$.

To determine the inelastic cross-section $\sigma_\nu$ for the
proton to interact with $\nu$ nucleons in \pA\ collisions, the standard Gribov
formalism~\cite{Bertocchi:1976bq} at high energies can be generalized
to include CF effects~\cite{Heiselberg:1991is}. When the impact
parameters in nucleon-nucleon ($NN$) interactions are small compared
to the typical distance between neighboring nucleons,

\begin{equation} 
  \begin{split}
    \quad 
    \sigma_{\nu} = & \int d\sigma P_{p}(\sigma)  \left(\begin{array}{c}
      A\\
      \nu \end{array} \right)
    \times \\
    & \int
    d\Vec{b}\, \left[{\sigma_{in}(\sigma) T(b)\over A}\right]^{\nu}
    \left[1-{\sigma_{in}(\sigma) T(b)\over A}\right]^{A-\nu}, 
  \end{split}
\end{equation} 
\noindent where $T(b) = \int_{-\infty}^{\infty}{dz} \rho(z,b)$ and
$\rho$ is the nuclear density distribution normalized such that
$\int\rho(r)\,d\Vec{r}=A$. $\sigma_{in}(\sigma)$ is the inelastic
cross-section for a configuration with the given total cross-section,
which following Ref.~\cite{Blaettel:1993rd,Alvioli:2014sba} is taken
to be a fixed fraction of $\sigma$. In the limit of no CF effects,
$P_p(\sigma)=\delta(\sigma-\left<\sigma\right>)$, and Eq.~4 
reduces to the Glauber model expression. The distribution over $\nu$
can be calculated with a Monte Carlo Glauber procedure, which includes
$NN$ correlations and finite size effects~\cite{Alvioli:2013vk}.
For $\nu \le 2 \left<\nu\right>$   the distribution over $\nu$  depends
mainly on $\omega_\sigma$ and only weakly on the exact form of
$P_p(\sigma)$~\cite{Alvioli:2013vk}. Although the Glauber
approximation ignores energy-momentum conservation in the inelastic
interaction of the proton with multiple nucleons, this does not modify
the calculation of $\sigma_\nu$,  or of the hadron multiplicity 
close to the nuclear fragmentation region~\cite{Alvioli:2014sba}.
(However, energy-momentum conservation effects may be important
in the evaluation of multiplicities at forward and central rapidities.)
This approach also accounts both for inelastic shadowing~\cite{Gribov}
and for the possibility of intermediate diffractive states between successive 
collisions~\cite{Blaettel:1993rd,Alvioli:2014sba}.

\begin{figure}[!t]
  \vskip -0.0cm
  \centerline{\includegraphics[width=8.0cm]{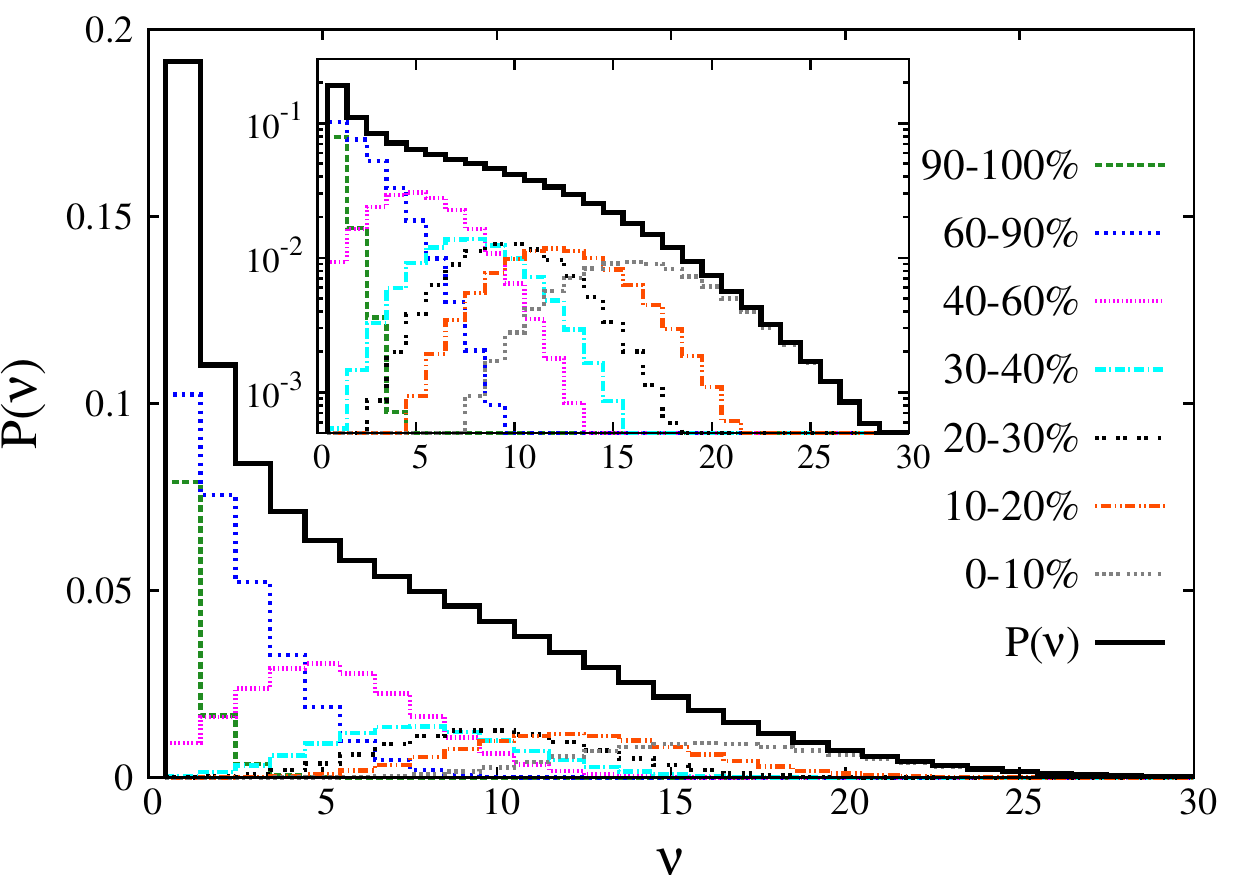}}
  \caption{(Color online) Probability distributions of $\nu$ proton-nucleon
    collisions in all \pA\ collisions and in those selected by different $\Sigma
    E_T$, or centrality, ranges. The inset shows the distributions on a
    log scale.}
  \label{fig1}
\end{figure}

ATLAS has studied the role of CF effects in interpreting the
correlation between hadron production at central rapidities and at
$-4.9 < \eta < -3.2$ in the nucleus-going direction in \pA\ collisions
at $\sqrt{s} = 5$~TeV~\cite{TheATLAScollaboration:2013cja}. The
total transverse energy, $\Sigma{E}_T$, near the nuclear fragmentation
region is not expected to be influenced by energy conservation effects
(due to the approximate Feynman scaling in this region), or to be strongly
correlated with the activity in the rapidity-separated central and
forward regions. This expectation is validated by a recent measurement
of $\Sigma{E}_T$ as a function of hard scattering kinematics in $pp$
collisions~\cite{ATLAS:2015hwg}. In Ref.~\cite{TheATLAScollaboration:2013cja},
$\Sigma{E}_T$ distributions were constructed as a function of the number of
participating nucleons $\nu + 1$.  Neglecting CF effects in calculating $\nu$
resulted in $\Sigma{E}_{T}$ distributions narrower than those observed in data.
Using the CF approach, which generally leads to a broader $\nu$ distribution from
the $\sigma > \left<\sigma\right>$ tail of $P_{p}(\sigma)$~\cite{Heiselberg:1991is},
with the value $\omega_{\sigma} = 0.1$ estimated for LHC energies
gives a reasonable description of the $\Sigma{E}_{T}$ data, while
$\omega_{\sigma}\sim 0.2 $ produces an overly broad distribution. The
resulting $\Sigma{E}_T(\nu)$ parameterization was used to calculate
the relative contributions from collisions with different $\nu$ values
to the \pA\ centrality classes (bins in $\Sigma E_T$) used by ATLAS.
Fig.~\ref{fig1} demonstrates that these centrality-selected
distributions have well separated mean values.

A challenging question is whether the fluctuations are modified or
amplified when a parton carrying a fraction $x$ of the projectile
momentum is present in the parton configuration, and, in particular,
whether large $x$ partons originate from configurations with smaller
than average $\sigma$. In the positronium example above, let the
$e^-(e^+)$ momentum in the rest frame be $\pm \vec{k}^{e}$, and apply
a boost to a fast frame. The light cone fraction carried by the
electron is $x_e=1/2 +k^{e}_3/2m_e$. If $\left|x_e-1/2\right| \gg
\left< \left|k_3^e\right|/2m_e\right>$, the positronium is squeezed,
leading to the cancellation of the photon field discussed above.

An analogous effect is present in the models of hadrons where few quarks
in the rest frame interact with each other through a potential which
is Coulomb-like at short distances, cf. the discussion in
section 5.1 of Ref.~\cite{Frankfurt:1994hf}. In these models,
the size of a given configuration decreases with increasing quark
momentum. Thus quarks with large $x$ in the fast frame arise from
configurations with large relative momenta in the rest frame and,
thus, a smaller size. The density of the gluon field in these
configurations is necessarily reduced.  Additionally, in QCD the main
contribution to the parton density in configurations with $x \gg
\left<x\right>$ is from configurations with the minimal number of
partons, leading to the quark counting rules~\cite{Brodsky:1973kr}. Hence for
such configurations, the nucleon's $q\bar q$ cloud is suppressed,
leading to a reduction of the soft cross-section even on the
non-perturbative scale~\cite{Frankfurt:1985cv}. This picture is corroborated
through a body of experimental evidence such as, for example,
measurements of coherent dijet production in pion--nucleus
collisions~\cite{Aitala:2000hc}.

Jet production measurements~\cite{ATLAS:2014cpa,Chatrchyan:2014hqa}
in inclusive \pA\ collisions at the LHC confirmed the pQCD expectations for
the total production rate. In the following, we focus on the data from ATLAS
which is directly adaptable to our analysis, although qualitatively similar effects
were observed by CMS \cite{Chatrchyan:2014hqa}. For centrality-selected \pA\
collisions, the jet production rate, $\sigma^{hard}_\nu$ for collisions with $\nu$
wounded  nucleons, was compared to the Glauber model expectation through the
ratio
\begin{eqnarray}
  R^{hard}_{\nu} = \left( \sigma^{hard}_\nu/\sigma_\nu \right ) /
  \left( \nu \cdot \sigma^{hard}_{NN}/\left<\sigma_{in}\right> \right),
  \label{hard}
\end{eqnarray}
where $\sigma^{hard}_{NN}$ is the jet rate in $NN $ collisions, and
$\left<\sigma_{in}\right>= \sigma^{NN}_{in}$. 
Large deviations from the expected $R^{hard}_\nu = 1$ were observed for
jets produced along the proton-going direction: namely, an enhancement
for peripheral (small $\nu$) collisions and a suppression for central
(large $\nu$) collisions, which compensate each other in the inclusive
cross section. These findings are not sensitive either to finite size
effects~\cite{Alvioli:2014sba} in the Glauber modeling nor, as
explained above, to energy-conservation effects. In the ATLAS
data~\cite{ATLAS:2014cpa}, $R^{hard}_\nu$ is presented as a function
of the fraction of the energy of the proton carried by the jet $z=E_{jet} /
E_p$, which, in the forward kinematics of interest, coincides with the
$x$ of the parton in the proton involved in the hard
interaction\footnote{Studying the correlation between the initial
  parton--parton kinematics and the final state kinematics of the
  produced jets in MC event generators confirms that events with a
  forward jet with energy $E$ have a distribution of $x$ values
  narrowly peaked around $x = 2 E / \sqrt{s}$.}. The data
demonstrate that for fixed energy release  in the nuclear hemisphere
($\Sigma{E}_{T}$) $R^{hard}_\nu$ is predominantly a function of $z$
and not of the jet $p_T$ or rapidity alone.

In our quantitative analysis, we combine the ATLAS model for the
$\nu$-dependence of $\Sigma E_T$ with the distribution of $\nu$ given
by a CF Monte Carlo approach \cite{Alvioli:2013vk,Alvioli:2014sba}.
The strength of the interaction $\sigma_{in}$ at given impact parameter,
$b$, is generated with the measure $P_{h}(\sigma)$ with inelastic profile
function $\Gamma_{in}(b)$ evaluated using the optical theorem. In the
evaluation, the elastic amplitude is taken to be proportional to
$\exp(Bt/2)$ with the $t$-slope $B$ fixed by the requirement that for
small $NN$ impact parameter the interaction is nearly black as it is
for the total $pp$ amplitude.  The transverse spread of partons in the
colliding nucleons were generated using generalized parton densities
which take into account a much stronger localization of hard
interactions relative to soft ones as well as the spatial $NN$
correlations in the nucleus~\cite{Alvioli:2009ab}. (For a detailed
description of the calculation of the hard collision rate as a
function of $\nu$ see Ref.~\cite{Alvioli:2014sba}.)

\begin{figure}[t]
  \vskip -0.0cm
  \centerline{\includegraphics[width=8.0cm]{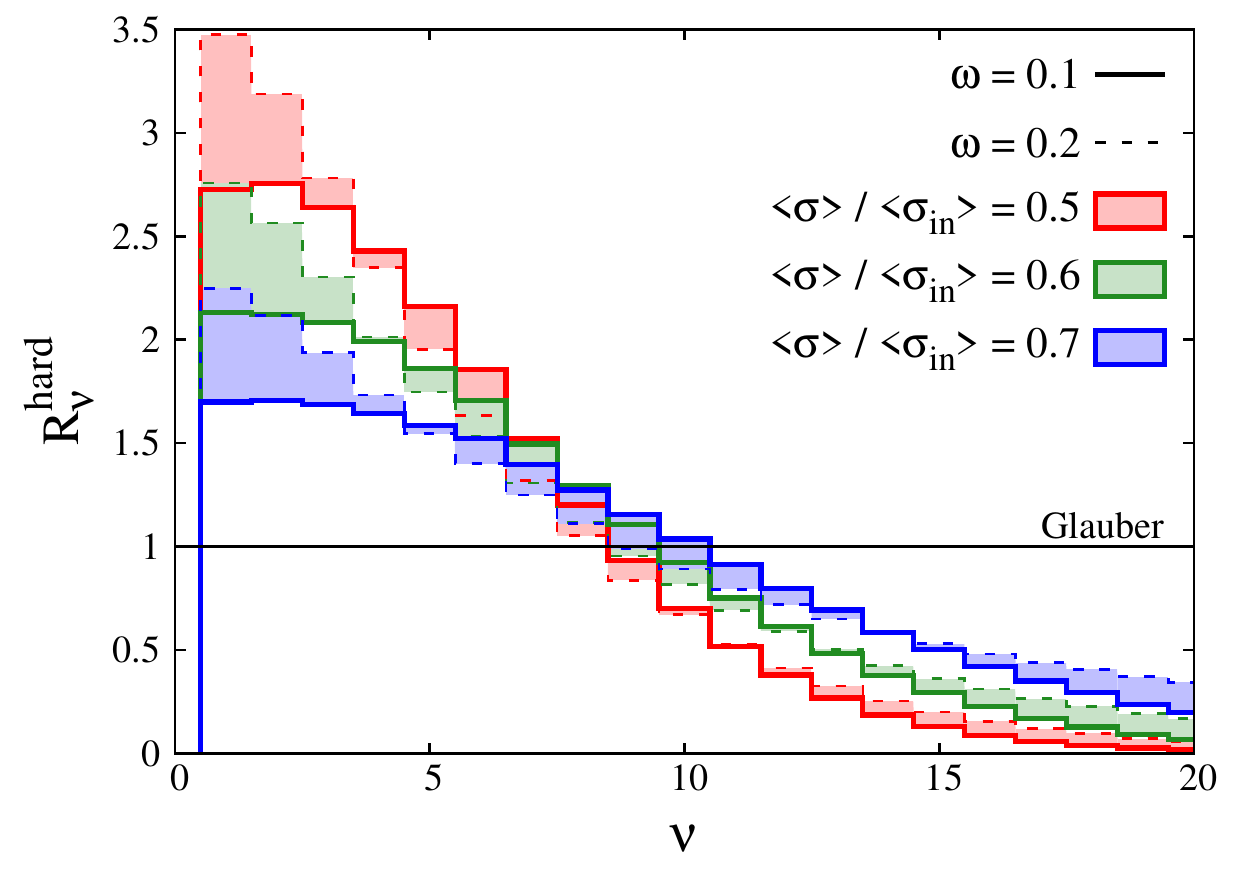}}
  \caption{(Color online) Probability of a hard process corresponding to a small
    $\lambda= \left<\sigma_{in}(x)\right>/\left<\sigma_{in}\right>$ trigger
    selection relative to that for a generic hard processes, as given in Eq.~\ref{hard}
    for $\omega_\sigma(x)=0.1, 0.2$. $R_\nu^{hard}=1$  is the expectation of the
    Glauber model.}
  \label{fig2}
\end{figure}

Fig.~\ref{fig2} shows $R^{hard}_\nu$ as a function of the ratio $\lambda$
of the average strength of the inelastic interaction for a trigger with a given
$x$, $\left<\sigma_{in}(x)\right>$ to that averaged over all
configurations, $\left<\sigma_{in}\right>$,
\begin{equation}
  \lambda= \left<\sigma_{in}(x)\right>/\left<\sigma_{in}\right>.
\end{equation}
\noindent
For the generic hard collisions we used Eq.~\ref{psigma}
with $\omega_\sigma=0.1$ which, within ATLAS, provides a good
description of soft production. For the small $\sigma_{in}(x)$ triggers in
Fig.~\ref{fig2}, we considered a range of $\lambda$
values and $\omega_\sigma(x)$ values between 0.1 and 0.2.
The relative jet production rate corresponding to small $\left<\sigma_{in}(x)\right>$
is enhanced at small $\nu$ and strongly suppressed at large $\nu$. With the exception
of the most peripheral and most central collisions, $R^{hard}_\nu$ depends primarily
on the mean $\left<\sigma_{in}(x)\right>$ and not on the choice of
$\omega_\sigma(x)$. Thus we fix $\omega_{\sigma}(x)=0.1$ and reduce the
dependence of $R^{hard}_\nu$ on $\nu$ to only one free $x$-dependent
parameter, $\lambda$.

\begin{figure}[t]
  \vskip -0.0cm
  \centerline{\includegraphics[width=8.0cm]{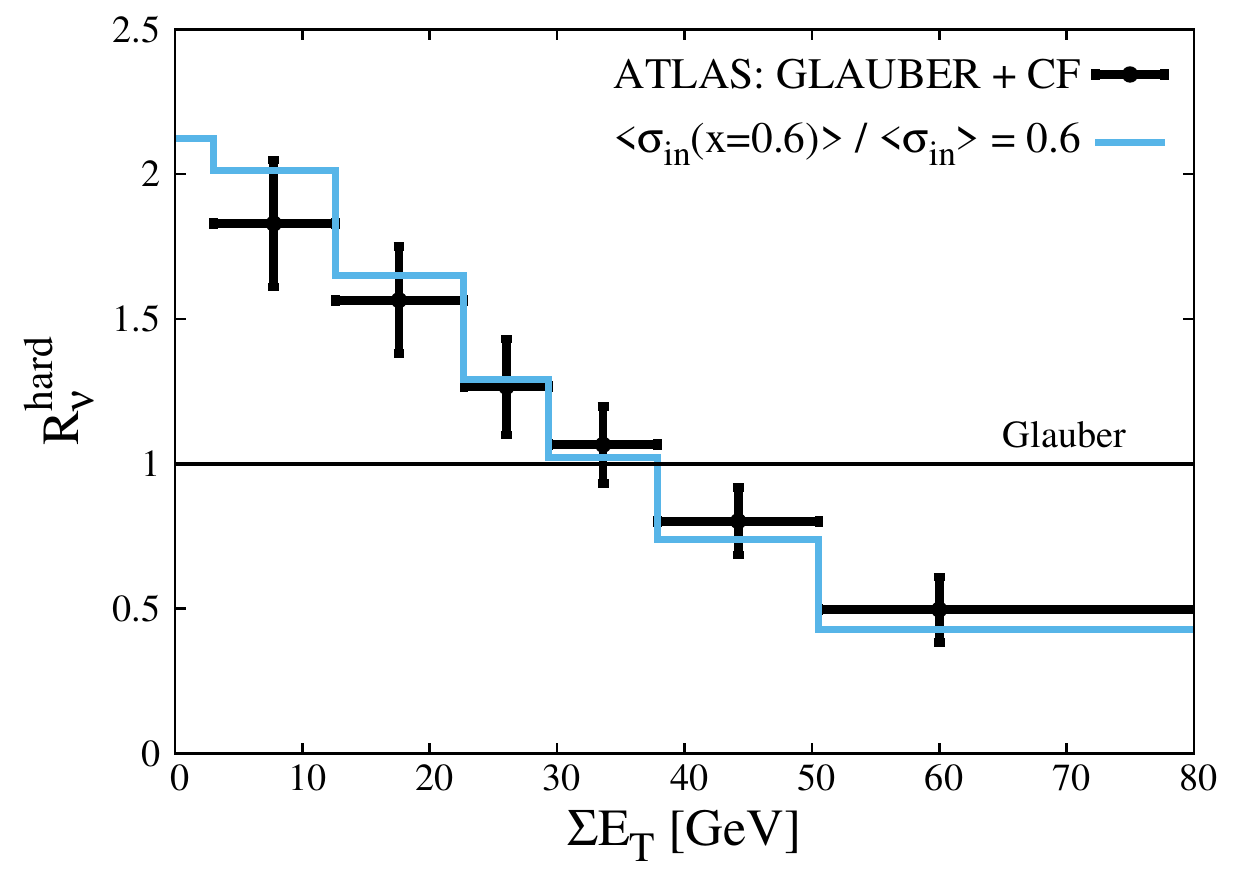}}
  \caption{(Color online) $R^{hard}_\nu$ for \pA\ collisions with $ x=
    E_{jet}/E_p=0.6$, and $\lambda=0.6$ for centrality bins extracted
    from the ATLAS data \cite{ATLAS:2014cpa} and using $\nu$ distributions
    given by the CF model~\cite{TheATLAScollaboration:2013cja}.  Errors
    are combined statistical and systematic errors.  The solid line is the Glauber
    model expectation.}
  \label{fig3}
\end{figure}

$R^{hard}_\nu$ values were then calculated for the specific centrality
bins used by ATLAS and compared to the $R^{hard}_\nu$ extracted from data.
For each centrality bin, we consider the measured
$R_{p\mathrm{Pb}}(p_\mathrm{T}\cosh{y})$ values from the four rapidity
intervals in the range $0.3 < y < 2.8$. These were fit
to a linear
function in $\log{\left(x/0.6\right)}$ in the range $0.04 < x < 1$,
with $x \equiv 2 p_\mathrm{T} \cosh{y} / \sqrt{s}$ and $y > 0$
denoting the proton-going direction, and the value at $x = 0.6$ was
extracted. Statistical uncertainties estimated by evaluating the RMS
deviation of the data points from the linear function in the region of
the fit were combined with systematic uncertainties on the data points
to yield total uncertainties. Figure~\ref{fig3} shows that
$\lambda \sim 0.6$ gives a good description of the data at $x =
0.6$. We emphasize that a naive interpretation of the data due to jet
energy loss cannot explain either the modification pattern in the
centrality-dependent $R^{hard}_\nu$, which features both enhancement
and suppression, or the observation of $R^{hard} = 1$ for inclusive
collisions, which follows from QCD factorization.

\begin{figure}[t]
  \centerline{\includegraphics[width=8.0cm]{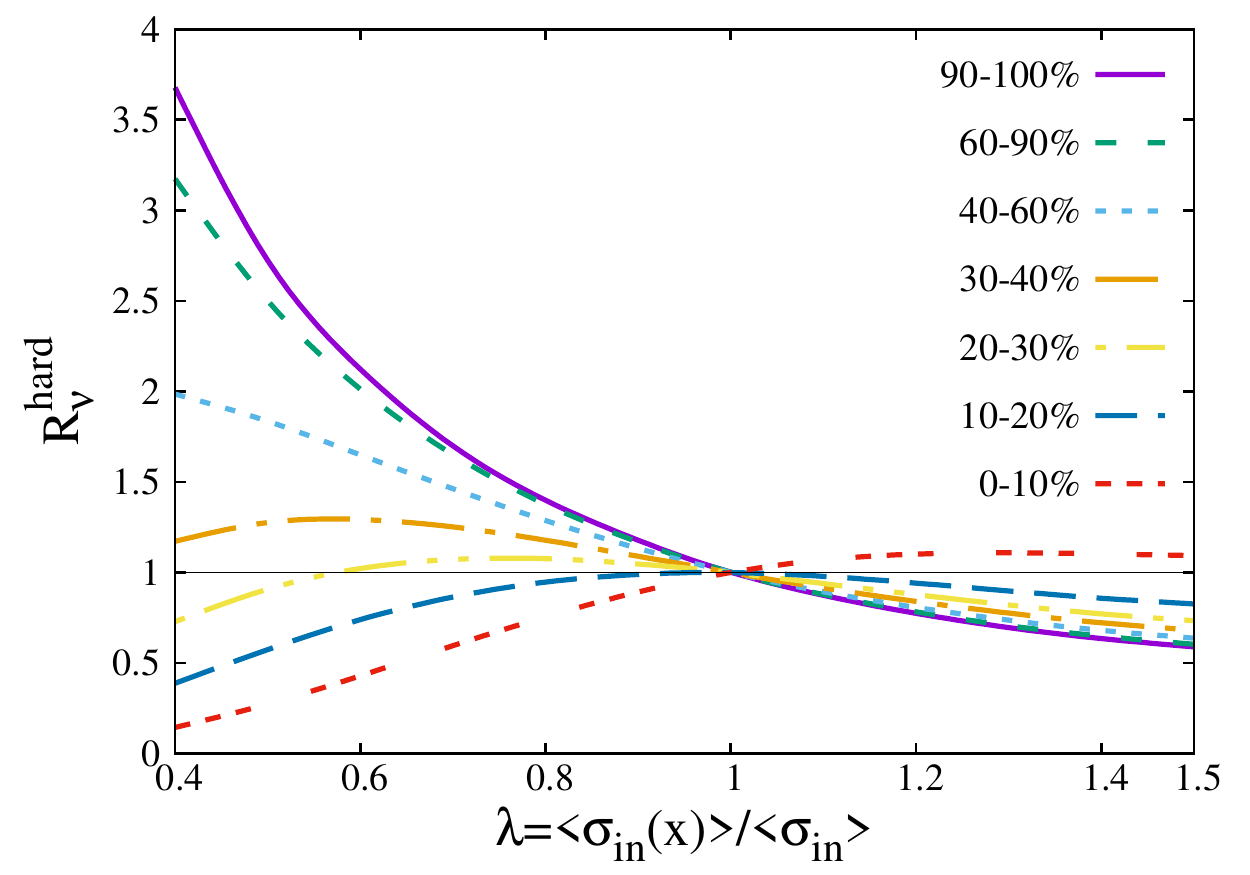}}
  \caption{(Color online) $R^{hard}_\nu$ for different centralities as a function of $\lambda$.}
  \label{fig4}
\end{figure}

Figure~\ref{fig4} shows the predictions of our model for $R^{hard}_\nu$
in each centrality bin as a function of $\lambda$. 
These predictions could be tested by extending the current analysis
of the LHC $pA$   data to $x< 0.6$ as well as by analyzing  the RHIC
$dA$ data~\cite{Adare:2015gla}.  The magnitude of the deviations
from $R^{hard}_\nu = 1$ increase smoothly with decreasing $\lambda < 1$.
For larger than average size configurations, corresponding to $\lambda > 1$,
the modification pattern reverses, producing an enhancement and suppression
in central and peripheral events respectively.

The agreement of the data for $x = 0.6$ with our calculation using
$\lambda = 0.6$ has a number of implications. It demonstrates that
large $x$ configurations have a weaker than average interaction
strength. More generally, it confirms the presence of CF effects in
\pA\ interactions and suggests that they should contribute to the
dynamics of central AA collisions~\cite{Heiselberg:1991is}.  It is in
line with the QCD quark counting rules which assume that large $x$
partons belong to configurations with a minimal number of constituents
interacting via hard gluon exchanges \cite{Brodsky:1973kr}. However, it is in
tension with approaches which neglect the short range correlations
between hadron constituents, such as the model in
\cite{Brodsky:2008pg}, and with those in which the transverse size of
the hadron is not squeezed at large $x$. 

To explore the energy dependence of this effect, the value of $\lambda$
at fixed $x$ can be determined at two different energies $\sqrt{s_1}$ and
$\sqrt{s_2}$ through the probability conservation of $P_h(\sigma)$:
$\int_0^{\sigma(\sqrt{s_1})} P_h(\sigma, \sqrt{s_1})d\sigma =
\int_0^{\sigma(\sqrt{s_2})} P_h(\sigma,
\sqrt{s_2})\ d\sigma$. At 30 GeV, $\lambda \approx 1/4$, a factor of
two smaller than at the LHC. This follows from the fact that in pQCD,
the cross-section of small-size configurations grows faster with
increasing collision energy than that of average configurations.

A weaker interaction strength for configurations with $x\ge 0.5$ has
implications for our understanding of the EMC effect.  This follows
from the analysis of Ref.~\cite{Frankfurt:1985cv}, in which the Schrodinger
equation for the bound state of the nucleus included a potential term
which depends on the internal coordinates of the nucleons. In this
potential, the overall attractive nature of the $NN$ interaction
results in a smaller binding energy for nucleons in small
configurations. Thus, by the variational principle, the probability of
such configurations is suppressed. The magnitude of this suppression
is comparable to the strength of the EMC effect at $ x\ge 0.5$, as
further discussed in Ref.~\cite{Freese:2014zda}.

Future \pA\ data in which the dijet kinematics are used to determine
$x$ and $x_\mathrm{A}$ on an event-by-event basis would allow for a
more detailed study of nucleon structure. In particular, studying the
modification pattern for hard processes to which gluons with $x\ge
0.3$ significantly contribute would probe whether the squeezing of the
transverse size is also present for configurations with large-$x$
gluons. Similarly, studying dijet production for $x\le 0.01$ would
allow for a study of configurations which have a higher than average
interaction strength. Measurements at lower energies would reveal the
energy dependence of $\left<\sigma(x)\right> $ and test whether it
grows much faster with energy than $\sigma$ as suggested by our
analysis above.  Finally, a comparison of $W^+, W^-$ production at
large $x_{p}$ would allow for a comparison of the transverse structure
of proton configurations with leading $u$ and $d$ quarks,
respectively.

{\bf Acknowledgments}
M.A.'s research was supported by grants provided by the Regione
Umbria, under contract POR-FESR Umbria 2007Ð2013, asse ii,
attivit\`a a1, azione 5, and by the Dipartimento della Protezione
Civile, Italy. B.A.C.'s research was supported by the US Department
of Energy Office of Science, Office of Nuclear Physics under Award
No. DE-FG02-86ER40281. L.F.'s research was supported by the
Binational Scientific Foundation Grant No. 0603216203. D.V.P.'s
research was supported by the US Department of Energy under
contract DE-AC02-98CH10886. M.S.'s research was supported by the
US Department of Energy Office of Science, Office of Nuclear Physics
under Award No.  DE-FG02-93ER40771.


\end{document}